\documentclass[preprint,showpacs,preprintnumbers,amsmath,amssymb]{revtex4}

\usepackage{graphicx}
\usepackage{dcolumn}
\usepackage{bm}
\usepackage{epsf}

\renewcommand{\bar}[1]{\overline{#1}}

\usepackage{amssymb}

\renewcommand{\bar}[1]{\overline{#1}}

\usepackage{indentfirst}
\usepackage{psfig,color}
\usepackage{epsfig}
\usepackage{epsf}
\usepackage{graphicx}

\begin{document}


\title{Nucleon sea in the effective chiral quark model}

\author{Yong Ding}
\author{Rong-Guang Xu}
\affiliation{Department of Physics, Peking University,
Beijing 100871, China}
\author{Bo-Qiang Ma}
\email{mabq@phy.pku.edu.cn} \altaffiliation{corresponding author.}
\affiliation{ CCAST (World Laboratory), P.O.~Box 8730, Beijing 100080, China\\
Department of Physics, Peking University, Beijing 100871,
China\footnote{Mailing address.}}

\begin{abstract}
The asymmetries of both light-flavor antiquark
$\bar{d}(x)-\bar{u}(x)$ and strange-antistrange $s(x)-\bar{s}(x)$
distributions of the nucleon sea are considered with more details
in the effective chiral quark model. We find that the asymmetric
distribution of light-flavor antiquarks $\bar{d}(x)-\bar{u}(x)$
matches the experiment data well and that the asymmetry of strange
and antistrange distributions can bring about 60-100\% correction
to the NuTeV anomaly of $\sin^{2}\theta_{w}$, which are three
standard deviations from the world average value measured in other
electroweak processes. The results on the correction to the NuTeV
anomaly are insensitive to the inputs of the constituent quark
distributions and the cut-off parameters. The ratios of
$\bar{d}(x)/\bar{u}(x)$ and $s(x)/\bar{s}(x)$ are also discussed,
and it is found that the ratio  $s(x)/\bar{s}(x)$ is compatible
with the available experiments with an additional symmetric sea
contribution being considered effectively.
\end{abstract}

\pacs{12.39.Fe, 11.30.Fs,  13.15.+g, 13.60.Hb}






\vfill



\vfill





\vfill 



\maketitle

\section{Introduction}
The physics community has been puzzled since the NuTeV
Collaboration reported the anomaly that the value:
$\sin^{2}\theta_{w}=0.2277\pm0.0013~(\mbox{stat})\pm0.0009~(\mbox{syst})$,
which was measured in deep inelastic scattering~(DIS) of
(anti-)neutrino from the nuclear target with possible
uncertainties and errors having been considered~\cite{zell02}, is
deviated from the standard model~(SM). The Weinberg angle (or weak
angle) $\theta_{w}$ is one of the important quantities in the
electroweak theory, and can be determined from various
experimental methods, such as atomic parity violation, $W$ and $Z$
masses, elastic and inelastic neutrino scattering, and so on. At
present, the world accepted value is
$\sin^{2}\theta_{w}=0.2227\pm0.0004$. The NuTeV Collaboration
measured the value of $\sin^{2}\theta_{w}$ by using the ratio of
(anti-)neutrino neutral-current to charged-current cross sections
on iron~\cite{zell02}, which is closely related to the
Paschos-Wolfenstein~(P-W) relation~\cite{pash73}:
\begin{equation}
R^{-}=\frac{\sigma^{{\nu}N}_{NC}-\sigma^{\overline{\nu}N}_{NC}}{\sigma^{{\nu}N}_{CC}
-\sigma^{\overline{\nu}N}_{CC}}=\frac{1}{2}-\sin^{2}\theta_{w}.
\label{ratio}
\end{equation}
The validity of this relation is based on the fundamental
assumptions of isoscalar target, charge symmetry and strange
symmetry $s(x)=\bar{s}(x)$, which give rise to a large number of
theoretical and experimental investigations on this issue.
Possible sources of the NuTeV anomaly beyond the SM have been
discussed~\cite{davidson02}. However, before speculating on the
possible new physics, one should first check carefully the {\it
standard} effect and the theoretical uncertainties within quantum
chromodynamics (QCD).

Here, we make a broad review of the main mechanisms which have
been used to investigate the NuTeV anomaly. The P-W relation holds
for a nuclear target provided that the nucleus is in the isoscalar
state, which means that various strong interaction effects must
cancel out in the ratio. But the actual targets used in the
neutrino experiments are usually non-isoscalar nuclei with a
significant neutron excess, such as the iron target in the NuTeV
experiment. Kumano~\cite{k02} investigated a conventional
explanation in terms of a nuclear modification caused by the
difference between the $u$- and $d$-valence distributions when the
charge and baryon-number conservation for nuclei are considered.
He and his collaborators estimated this effect to the NuTeV
$\sin^{2}\theta_{w}$ value by using a $\chi^{2}$ analysis method
to reproduce nuclear data on the structure function $F_{2}$ and
the cross-section of Drell-Yan processes, and noticed that the
effect is not large enough to explain the whole NuTeV anomaly in
their later work~\cite{k0412307}. And Kulagin~\cite{kulagin03} not
only took into account the neutron excess correction to the P-W
relation for a neutron-rich target, but also discussed other
corrections to the P-W relation caused by other nuclear effects:
the Fermi motion, nuclear binding, nuclear shadowing, etc. In
addition, nuclear effects were completely estimated and the shift
to the NuTeV anomaly was given with some uncertainties in
Ref.~\cite{ksy02} by using a particular nuclear $x$-rescaling
model to describe the structure functions in
(anti-)neutrino-nucleus DIS. Also, there are other
suggestions~\cite{0204007} from a conservative point of views.

Another fundamental assumption is charge symmetry which is a more
restricted form of isospin invariance involving a rotation of
$180^{\circ}$ about the ``2" axis in isospin space, or more
specifically, the isospin symmetry for the $u \leftrightarrow d$
exchanges between protons and neutrons~\cite{ma92}. However, there
have been some discussions about the correction to the P-W
relation due to this symmetry breaking. Sather~\cite{sather92}
first pointed out that the charge symmetry violation~(CSV) should
largely affect the extraction of $\sin^{2}\theta_{w}$ from the
neutrino collision and gave an estimation about it. The similar
result was given in Ref.~\cite{lt03} that CSV effect should reduce
about one-third of the discrepancy between the NuTeV result and
the accepted average value of $\sin^{2}\theta_{w}$. On the
contrary, Cao and Signal~\cite{cs03} calculated the
nonperturbative effect of CSV within the meson cloud model
framework and showed no contribution to the NuTeV anomaly. At this
stage, we still have no direct experimental evidences to point to
a substantial violation of CS in parton distributions. Only there
are some up limits in experiment for the CSV, because the effect
of CSV usually confuses with the asymmetry of $s$ and $\bar{s}$
distributions.

The asymmetry of strange and antistrange quark distributions may
imply an additional source for the anomaly.
The topic of strange content in the nucleon sea is one of the most
challenging issues in the hadron physics, especially for its
connection to the proton spin problem and to the NuTeV anomaly.
For the part of nucleon sea generated through gluon splitting
$g\rightarrow q\bar{q}$ in perturbative QCD, CP symmetry is
expected in the quark and antiquark distributions in the leading
order. It has been argued recently that there is a
strange-antistrange asymmetry in perturbative QCD at
three-loop~\cite{pQCD}, however, the perturbative QCD alone
definitely predicts a non-vanishing and $Q$-dependent value of
strange-antistrange asymmetry which would have a very small
contribution to the extraction of $\sin^{2}\theta_{w}$. So the
cause of a sizable asymmetry of the nucleon strange sea should be
of nonperturbative origin~\cite{bm96,ST,bw}. Furthermore, it is
highly probable that the momentum distributions of strange and
antistrange are different in the nucleon sea, although the total
number of strange quark and antiquark occurring as quantum
fluctuations must be precisely equal to conserve the strangeness
quantum number. The consequence on the NuTeV anomaly from possible
asymmetry of strange and antistrange quark distributions in the
nucleon sea was examined by Cao and Signal~\cite{cs03} utilizing
the meson cloud model~\cite{ST,hss96} and the result is fairly
small and has no significant effect on the NuTeV result.
Oppositely, it was shown~\cite{dm04} that the light-cone
baryon-meson fluctuation model proposed by Brodsky and
Ma~\cite{bm96} can describe the second moment
$S^{-}\equiv\int^{1}_{0} x[s(x)-\bar{s}(x)]\textmd{d}x$ and
produce an asymmetry of $s(x)$ and $\bar{s}(x)$ distributions
which could remove roughly 30-80\% of the discrepancy between the
NuTeV result and other determination of $\sin^{2}\theta_{w}$.
Also, in Ref.~\cite{alwall}, the asymmetric $s-\bar{s}$
distribution treated in a different framework of non-perturbative
hadronic $K+\Lambda$ fluctuations is found to reduce the NuTeV
result to only about 2 standard deviations from SM, while the
realistic behavior of the strange sea distribution $ [x s(x)+x
\bar{s}(x)]/2$ is also reasonably reproduced. Kretzer
$et~al.$~\cite{kretzer04} obtained QCD correction by calculating
the next-to-leading-order~(NLO) neutrino cross sections and
studied the shift of $\sin^{2}\theta_{w}$ which is closely
correlated with $s(x)\neq\bar{s}(x)$ parton distribution function
sets and isospin violation with some uncertainties. It is also
noticed that the contribution to the NuTeV anomaly changes sign
from ``$-$" to ``$+$" when $S^{-}$ is calculated~\cite{olness03}
by using the Lagrange Multiplier method. A more reliable
theoretical analysis in the chiral quark model has been give in
Ref.~\cite{dxm04}, in which it is shown that the NuTeV anomaly can
be accounted for by at least 60\% without sensitivity to the
inputs of constituent quark distributions and cut-off parameters.
Recently, Wakamatsu~\cite{wakamatsu04} gave also a theoretical
analysis based on the flavor SU(3) chiral quark soliton model by
introducing a parameter of the effective mass difference between
strange and nonstrange quarks and predicted a fairly large
asymmetry of $s$- and $\bar{s}$-quark distributions which would
solely resolve the NuTeV anomaly, in similar to the conclusion in
Ref.~\cite{dxm04}.
In addition, Szczurek $et~al.$~\cite{sbf96} discussed more earlier
the dressing of constituent quark with a pseudoscalar meson cloud
within the effective chiral quark model by including the effect of
SU(3)$_{f}$ symmetry violation explicitly. They found more pions
and kaons in the nucleon than what predicted by traditional meson
cloud model and pointed out that the effective chiral quark model
may lead to a sizeable asymmetric strangeness content of the
nucleon sea.

As we know that the nucleon sea has received attention for a long
time because of the abundant phenomena which are away from the
naive theoretical expectations. For example, the Gottfried sum
rule~(GSR)~\cite{g67} violation is related to the light-flavor
antiquark asymmetry in the nucleon sea and has been measured in
several experiments. At the same time, there are some models, such
as: the meson cloud model~\cite{mc01,c99} in which the relative
success was obtained in both the difference and the ratio of
$\bar{d}$ and $\bar{u}$ by including the perturbative component;
Pauli blocking effect which first suggested by Field and
Feynman~\cite{ff77}; and several different types of chiral models
and so on, proposed to explain the violation of Gottfried sum
rule. And another case is the proton spin crisis~\cite{a89} and
its connection to the strange content in the nucleon
sea~\cite{bek88}. Usually, the distributions of $s$ and $\bar{s}$
are assumed to be symmetric, but this is neither proved
theoretically nor experimentally. Moreover, some measurements and
analyses~\cite{b95,sbr97,a97} show that the distributions of
strange and antistrange quark may be asymmetric. A joint
fit~\cite{bpz00} to the neutrino charged-current cross
sections~\cite{b91} and charged lepton structure function data was
made by the CDHS Collaboration and some improvements were achieved
in the fits if asymmetry in the strange sea was allowed.

In this paper, we present the more detailed calculation of the
asymmetric distributions for both the light-flavor antiquarks
$\bar{d}(x)$ and $\bar{u}(x)$ and the strange content in the
nucleon sea by using the effective chiral quark model along with
our previous work~\cite{dxm04}, and find that the distributions of
$\bar{d}(x)-\bar{u}(x)$, $\bar{d}(x)+\bar{u}(x)$,
$x(s(x)+\bar{s}(x))$ and $s(x)/\bar{s}(x)$ match well with the
experimental data, when additional symmetric sea contributions
being considered effectively by taking into account the difference
between model results and data parametrization. Furthermore, it is
should be pointed out that the asymmetry of $s(x)$ and
$\bar{s}(x)$ distributions could remove roughly the NuTeV anomaly
by at least 60\%, and it is more remarkable that this conclusion
is insensitive to the different inputs of the effective chiral
quark model.

\section{The sea content in the effective chiral quark model\label{Sec:Two}}

The effective chiral quark model, established by
Weinberg~\cite{Weinberg}, and developed by Manohar and
Georgi~\cite{mg84}, has been widely accepted by the hadron physics
society as an effective theory of QCD at low energy scale. The
effective chiral quark model has an apt description of its
important degrees of freedom in terms of quarks, gluons and
Goldstone~(GS) bosons at momentum scales relating to hadron
structure. There has been a prevailing impression that the
effective chiral quark model is successful in explaining the
violation of GSR (which was first detected by the New Muon
Collaboration at CERN~\cite{nmc91}) by the analysis of Eichten,
Hinchliffe and Quigg~\cite{ehq92}. Also, this model plays an
important role in explaining the proton spin crisis~\cite{a89} by
Cheng and Li~\cite{cl95}. A recent study by us~\cite{dxm04} shows
that the strange-antistrange asymmetry within the effective chiral
quark model could explain the NuTeV anomaly also. It is the
purpose of this paper to provide a more detailed analysis on the
strange-antistrange asymmetry and its connection to the
light-flavor sea asymmetry in the effective chiral quark model.

\begin{figure}[!htbp]
\begin{center}
\scalebox{0.7}{\includegraphics{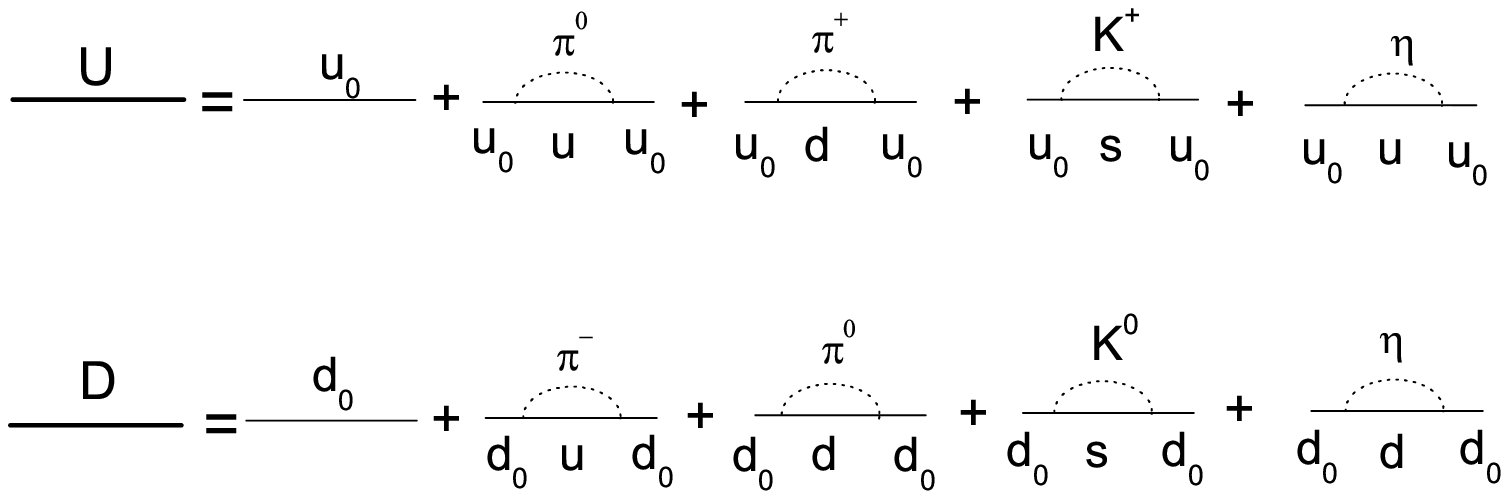}}
 \caption{The dressing of
the constituent U and D quarks with Goldstone
bosons.}\label{wavefunction}
\end{center}
\end{figure}

\begin{figure}[!htbp]
\begin{center}
\scalebox{0.8}{\includegraphics{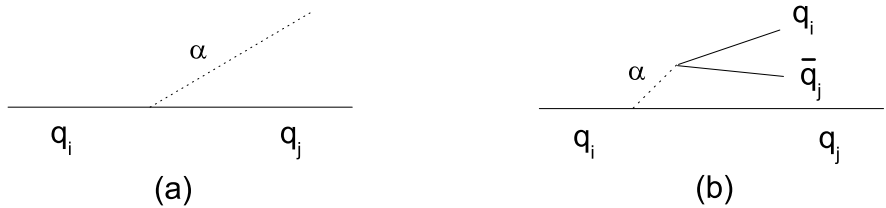}}\caption{\small (a) A
constituent quark $q_{i}$  fluctuates  into a Goldstone boson
$\alpha$ plus a recoil constituent quark $q_{j}$. (b) A Goldstone
boson emits a $q_{i}$ and a $\bar{q}_{j}$ .}\label{fluctuation}
\end{center}
\end{figure}

The chiral symmetry at high energy scale and its breaking at low
energy scale are the basic properties of QCD. Because the effect
of the internal gluons is small in the effective chiral quark
model at low energy scale, the gluonic degrees of freedom are
negligible when comparing to those of the GS bosons and quarks. In
this picture, the valence quarks which are contained in the
nucleon fluctuate into quarks plus GS bosons, which spontaneously
break chiral symmetry, and any low energy hadron properties should
include this symmetry violation. The effective interaction
Lagrangian is
\begin{equation}
L=\bar{\psi}(iD_{\mu}+V_{\mu})\gamma^{\mu}\psi+ig_{A}\bar{\psi}A_{\mu}\gamma^{\mu}\gamma_{5}\psi+\cdots,
\end{equation}
where
\begin{equation}
\psi=\left(%
\begin{array}{c}
  u \\
  d \\
  s \\
\end{array}%
\right)
\end{equation}
is the quark field and $D_{\mu}=\partial_{\mu}+igG_{\mu}$ is the
gauge-covariant derivative of QCD, with $G_{\mu}$ standing for the
gluon field, $g$ standing for the strong coupling constant and
$g_{A}$ for the axial-vector coupling constant determined from the
axial charge of the nucleon. $V_{\mu}$ and $A_{\mu}$ are the
vector and axial-vector currents which are defined by the
following forms:
\begin{equation}
\left(%
\begin{array}{c}
  V_{\mu} \\
  A_{\mu} \\
\end{array}%
\right)=\frac{1}{2}(\xi^{+}\partial_{\mu}\xi\pm\xi\partial_{\mu}\xi^{+}),
\end{equation}
where $\xi=\mathrm{exp}(i\Pi/f)$, and $\Pi$  has the form:
\begin{equation}
\Pi\equiv\frac{1}{\sqrt{2}}\left(
\begin{array}{ccc}
  \frac{\pi^{0}}{\sqrt{2}}+\frac{\eta}{\sqrt{6}} & \pi^{+} & K^{+} \\
  \pi^{-} & -\frac{\pi^{0}}{\sqrt{2}}+\frac{\eta}{\sqrt{6}} & K^{0} \\
  K^{-} & \bar{K^{0}} & \frac{-2\eta}{\sqrt{6}} \\
\end{array}
\right).
\end{equation}
With the expansion for $V_{\mu}$ and $A_{\mu}$ in powers of
$\Pi/f$, it gives $V_{\mu}=0+O(\Pi/f)^{2}$ and
$A_{\mu}=i\partial_{\mu}\Pi/f+O(\Pi/f)^{2}$, where the
pseudoscalar decay constant is $f\simeq93$~MeV. So the effective
interaction Lagrangian between GS bosons and quarks in the leading
order becomes~\cite{ehq92}
\begin{equation}
L_{\Pi
q}=-\frac{g_{A}}{f}\bar{\psi}\partial_{\mu}\Pi\gamma^{\mu}\gamma_{5}\psi.
\end{equation}
The framework that we use in this paper is based on the
time-ordered perturbative theory in the infinite momentum
frame~(IMF). Because all particles are on-mass-shell in this frame
and the factorization of the subprocess is automatic, we neglect
all possible off-mass-shell corrections. In this framework, we can
express the quark distributions inside a nucleon as a convolution
of a constituent quark distribution in a nucleon and the structure
functions of a constituent quark. The light-front Fock
decompositions of constituent quark wave functions have the
following forms
\begin{equation}
|U\rangle=\sqrt{Z}|u_{0}\rangle+a_{\pi}|d\pi^{+}\rangle+\frac{a_{\pi}}{%
\sqrt{2}}|u\pi^{0}\rangle+a_{K}|sK^{+}\rangle+\frac{a_{\eta}}{\sqrt{6}}%
|u\eta\rangle,\label{u}
\end{equation}
\begin{equation}
|D\rangle=\sqrt{Z}|d_{0}\rangle+a_{\pi}|u\pi^{-}\rangle+\frac{a_{\pi}}{%
\sqrt{2}}|d\pi^{0}\rangle+a_{K}|sK^{0}\rangle+\frac{a_{\eta}}{\sqrt{6}}%
|d\eta\rangle,\label{d}
\end{equation}
and the corresponding picture are shown in
Fig.~\ref{wavefunction}.
Here, $Z$ is the renormalization constant for the bare constituent
quarks which are massive and denoted by $|u_{0}\rangle$ and
$|d_{0}\rangle$, and $|a_{\alpha}|^{2}$ are the probabilities to
find GS bosons in the dressed constituent quark states $|U\rangle$
for an $up$ quark and $|D\rangle$ for a $down$ quark, where
$\alpha=\pi, K, \eta$. In the effective chiral quark model, the
fluctuation of a bare constituent quark into a GS boson and a
recoil bare constituent quark can be given~\cite{sw98}:
\begin{equation}
q_{j}(x)=\int^{1}_{0}\frac{\textmd{d}y}{y}P_{j\alpha/i}(y)q_{i}(\frac{x}{y}).\label{q}
\end{equation}
This process also can be described in Fig.~\ref{fluctuation}(a).
In Eq.~(\ref{q}), $P_{j\alpha/i}(y)$ is the splitting function
which gives the probability for finding a constituent quark $j$
carrying the light-cone momentum fraction $y$ together with a
spectator GS boson~$\alpha$ and having the following form:
\begin{eqnarray}
P_{j\alpha/i}(y)=\frac{1}{8\pi^{2}}(\frac{g_{A}\bar{m}}{f})^{2}\int
\textmd{d}k^{2}_{T}\frac{(m_{j}-m_{i}y)^{2}+k^{2}_{T}}{y^{2}(1-y)[m_{i}^{2}-M^{2}_{j\alpha}]^{2}},
\label{splitting}
\end{eqnarray}
where $m_{i}, m_{j}, m_{\alpha}$ are the masses of the $i,
j$-constituent quarks and the pseudoscalar meson $\alpha$,
respectively, $\bar{m}=(m_{i}+m_{j})/2$ is the average mass of the
constituent quarks, and $M_{j\alpha}^{2}$ is the invariant mass
square of the final states:
\begin{equation}
M^{2}_{j\alpha}=\frac{m^{2}_{j}+k^{2}_{T}}{y}+\frac{m^{2}_{\alpha}+k^{2}_{T}}{1-y}.
\end{equation}
In this paper, we also adopt the definition of the  moment of
splitting function:
\begin{equation}
\langle x^{n-1}
P_{j\alpha/i}\rangle=\int^{1}_{0}x^{n-1}P_{j\alpha/i}(x)\textmd{d}x
\end{equation}
with the first moment $\langle P_{j\alpha/i}\rangle=\langle
P_{\alpha j/i}\rangle\equiv \langle
P_{\alpha}\rangle=|a_{\alpha}|^{2}$~\cite{sw98}. It is
conventional that an exponential cutoff is used in IMF
calculations because the integral in Eq.~(\ref{splitting})
requires a momentum cutoff function at the quark-GS boson vertex.
Commonly,
\begin{equation}
g_{A}\rightarrow
g_{A}^{\prime}\textmd{exp}\bigg{[}\frac{m^{2}_{i}-M^{2}_{j\alpha}}{4\Lambda^{2}}\bigg{]},
\end{equation}
with $g_{A}^{\prime}=1$ following the large $N_{c}$
argument~\cite{w90}, however, $g_{A}^{\prime}=0.75$ was taken in
the original work~\cite{mg84}. And $\Lambda$ is the cutoff
parameter, which is determined by the experimental data of the
Gottfried sum and the constituent quark mass inputs for the pion.
The experimental value of Gottfried sum is
$0.235\pm0.0026$~\cite{nmc91}, which is different from the value
1/3 predicted by the naive parton model in which both the
light-flavor sea and the isospin between proton and neutron are
SU(2) symmetric~\cite{ma92}. Employing the quark distributions of
the effective chiral quark model, we get the Gottfried sum
determined by the difference between proton and neutron structure
functions:
\begin{eqnarray}
S_{\mathrm{Gottfried}}&=&\int^{1}_{0}\frac{\mathrm{d}x}{x}[F^{p}_{2}(x)-F^{n}_{2}(x)]\nonumber\\
&=&\frac{1}{3}\int_{0}^{1}\frac{dx}{x}[u(x)+\bar{u}(x)-d(x)-\bar{d}(x)]\nonumber\\&
=&\frac{1}{3}(Z-\frac{1}{2}\left<P_{\pi}\right>+\left<P_{K}\right>+\frac{1}{6}\left<P_{\eta}\right>)\nonumber\\
&=&\frac{1}{3}(1-2\left<P_{\pi}\right>).\label{gottfried}
\end{eqnarray}
From above equation, we gain the values of Gottfried sum given in
table~\ref{value} corresponding to different $\Lambda_{\pi}$, from
which we can find that the appropriate value for $\Lambda_{\pi}$
is 1500~MeV, when the value of Gottfried sum  matches well with
the experimental data. At the same time, $\langle
P_{\pi}\rangle=0.149, \langle P_{K}\rangle=0.085, \langle
P_{\eta}\rangle=0.063$ and $Z=0.682$. But for $K$ and $\eta$
mesons, the terms $\langle P_{K}\rangle$ and $\langle
P_{\eta}\rangle$ in the Gottfried sum cancel out those terms in
$Z=1-\frac{3}{2}\langle P_{\pi}\rangle-\langle
P_{K}\rangle-\frac{1}{6}\langle P_{\eta}\rangle$, so the value of
$\Lambda$ can not be determined from Eq.~(\ref{gottfried}) or
experimental data. It is natural to assume that the cutoffs are
same for $\pi$, $K$ and $\eta$ mesons in the effective chiral
quark model:
$\Lambda_{\pi}=\Lambda_{K}=1500$~MeV~\cite{sbf96,sw98}, which is
different from the traditional meson cloud model.

\begin{figure}
\begin{center}
\scalebox{0.98}{\includegraphics[0,16][310,230]{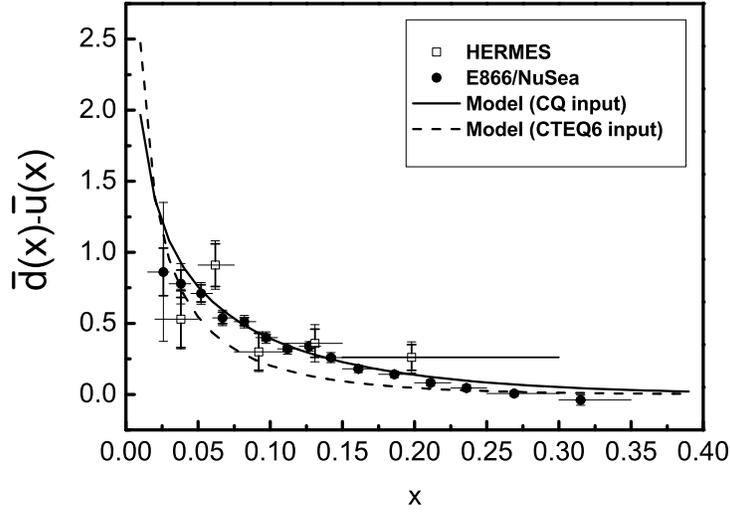}}
\caption{\small Distributions for $\bar{d}(x)-\bar{u}(x)$ for
$\Lambda_{\pi}=1500$~MeV. The solid and dashed curves are the
model calculation results in the effective chiral quark model with
the constituent quark (CQ) model and the CTEQ6 parametrization as
inputs for $u_0(x)$ and $d_0(x)$, respectively. The data are from
HERMES~($Q^{2}=2.3$~GeV$^{2}$)~\cite{h98} and
E866/NuSea~($Q^{2}=54$~GeV$^{2}$) experiments~\cite{e01}.
}\label{udx}
\end{center}
\end{figure}

\begin{figure}
\begin{center}
\scalebox{0.98}{\includegraphics[0,16][310,230]{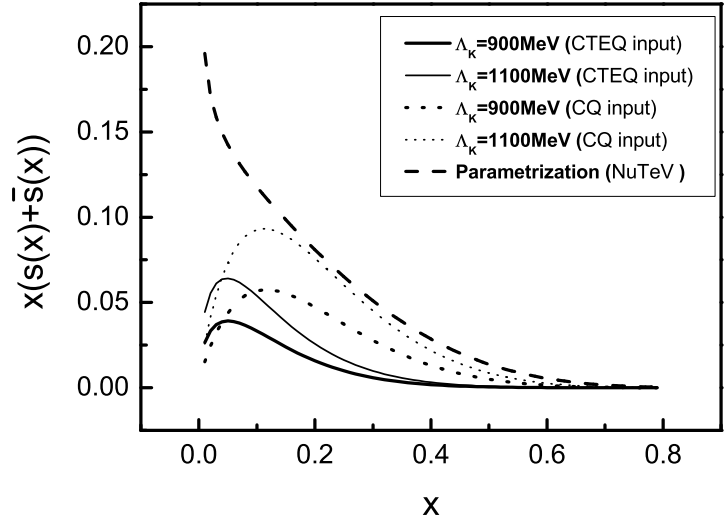}}
\caption{\small Distributions for $x(s(x)+\bar{s}(x))$.
 The solid~(dotted) curves are the calculated results in the effective chiral quark model with the CTEQ6
parametrization~(CQ model) as input. The thick solid~(dotted)
curves are for $\Lambda_{K}=900$~MeV and the thin ones are for
$\Lambda_{K}=1100$~MeV, respectively. The dashed curve is for the
parametrization of NuTeV data~\cite{NUTEV}. }\label{xsplussbar}
\end{center}
\end{figure}

\begin{table} [!htbp]\caption {\label{value}The values for  $\Lambda$ and Gottfried sum}
\begin{center}
\begin{tabular}{|c|c|c|c|c|c|c|c|}
  \hline
  \hline
  $\Lambda_{\pi}$ & 1000 & 1100 & 1200 & 1300 & 1400 & 1500 & 1600 \\
  \hline
     Gottfried sum    & 0.268 & 0.261 & 0.253 & 0.247 & 0.240 & 0.234 & 0.223 \\
    \hline
\end{tabular}
\end{center}
\end{table}
When probing the internal structure of GS bosons in
Fig.~\ref{fluctuation}(b), we can also write the process in the
following form~\cite{sw98}:
\begin{equation}
q_{k}(x)=\int\frac{\textmd{d}y_{1}}{y_{1}}\frac{\textmd{d}
y_{2}}{y_{2}}V_{k/\alpha}(\frac{x}{y_{1}})P_{\alpha
j/i}(\frac{y_{1}}{y_{2}})q_{i}(y_{2}),
\end{equation}
where $P_{\alpha j/i}(x)=P_{j\alpha/i}(1-x)$,  $V_{k/\alpha}(x)$
is the quark $k$ distribution function in $\alpha$ and satisfies
the normalization $\int_{0}^{1}V_{k/\alpha}(x)dx=1$. Because the
mass of $\eta$ is so high and the coefficient is so small so that
the fluctuation of it is suppressed, the contribution is not
considered in our calculation in this paper. From Eqs.~(\ref{u})
and (\ref{d}), we can have the quark distribution functions of the
nucleon by using the splitting function Eq.~(\ref{splitting}) and
the constituent quark distributions $u_{0}$ and $d_{0}$ which are
normalized to 1,
\begin{eqnarray}
u(x)&=&Zu_{0}(x)+P_{u\pi^{-}/d}\otimes d_{0}+V_{u/\pi^{+}}\otimes
P_{\pi^{+}d/u}\otimes
u_{0}+\frac{1}{2}P_{u\pi^{0}/u}\otimes u_{0}\nonumber\\&+&V_{u/K^{+}}\otimes P_{K^{+}s/u}\otimes u_{0}+
\frac{1}{4}V_{u/\pi^{0}}\otimes (P_{\pi^{0}u/u}\otimes u_{0}+P_{\pi^{0}d/d}\otimes d_{0}),\nonumber\\
d(x)&=&Zd_{0}(x)+P_{d\pi^{+}/u}\otimes u_{0}+V_{d/\pi^{-}}\otimes
P_{\pi^{-}u/d }\otimes d_{0}+ \frac{1}{2}P_{d\pi^{0}/d}\otimes
d_{0}\nonumber\\&+&V_{d/K^{0}}\otimes P_{K^{0}s/d}\otimes d_{0}
+\frac{1}{4}V_{d/\pi^{0}}\otimes (P_{\pi^{0}u/u }\otimes
u_{0}+P_{\pi^{0}d/d}\otimes d_{0}).~\label{uddis}
\end{eqnarray}
Here, we define the notation for the convolution integral:
\begin{equation}
P\otimes q=\int_{x}^{1}\frac{\textmd{d}y}{y}P(y)q(\frac{x}{y}),
\end{equation}
and
\begin{equation}
V\otimes P\otimes
q=\int_{x}^{1}\frac{\textmd{d}y_{1}}{y_{1}}\int_{y_{1}}^{1}\frac{\textmd{d}y_{2}}{y_{2}}V(
\frac{x}{y_{1}})P(1-\frac{y_{1}}{y_{2}})q_{i}(y_{2}).
\end{equation}
In the same way, we can have the light-flavor antiquark and
strange quark and antiquark distributions:
\begin{eqnarray}
\bar{u}(x)&=&V_{\bar{u}/\pi^{-}}\otimes P_{\pi^{-}u/d}\otimes
d_{0}+\frac{1}{4}V_{\bar{u}/\pi^{0}}\otimes (P_{\pi^{0}u/u}\otimes
u_{0}+P_{\pi^{0}d/d}\otimes d_{0}),\nonumber\\
\bar{d}(x)&=&V_{\bar{d}/\pi^{+}}\otimes P_{\pi^{+}d/u}\otimes
u_{0}+\frac{1}{4}V_{\bar{d}/\pi^{0}}\otimes (P_{\pi^{0}u/u}\otimes
u_{0}+P_{\pi^{0}d/d}\otimes d_{0}),\nonumber\\
 s(x)&=&P_{sK^{+}/u}\otimes u_{0}+P_{sK^{0}/d}\otimes d_{0},\nonumber\\
\bar{s}(x)&=&V_{\bar{s}/K^{+}}\otimes P_{K^{+}s/u}\otimes
 u_{0}+V_{\bar{s}/K^{0}}\otimes P_{K^{0}s/d}\otimes d_{0},\label{udsdis}
\end{eqnarray}
where \begin{eqnarray}
&&~~~V_{u/\pi^{+}}=V_{\bar{d}/\pi^{+}}=V_{d/\pi^{-}}=V_{\bar{u}/\pi^{-}}\nonumber\\&&=2V_{u/\pi^{0}}
=2V_{\bar{u}/\pi^{0}}=2V_{d/\pi^{0}}=2V_{\bar{d}/\pi^{0}}
\nonumber\\&&=\frac{1}{2}V_{\pi}(x,\mu_{\mathrm{NLO}}^{2}),
\end{eqnarray} and
$$V_{\bar{s}/K^{+}}=V_{\bar{s}/K^{0}}, V_{u/K^{+}}=V_{d/K^{0}}.$$
From above equations, we can reexamine the valence quark
distributions $u_{v}(x)=u(x)-\bar{u}(x)$ and
$d_{v}(x)=d(x)-\bar{d}(x)$ which satisfy the correction
normalization with the renormalization constant $Z$. The
parameterizations of parton distributions for mesons are taken
from GRS98~\cite{grs98} and the sea content in the nucleon is not
considered here,
\begin{equation}
V_{\pi}(x,\mu_{\mathrm{NLO}}^{2})=1.052x^{-0.495}(1+0.357\sqrt{x})(1-x)^{0.365},
\end{equation}
\begin{equation}
V_{u/K^{+}}(x,\mu_{\mathrm{NLO}}^{2})=V_{d/K^{0}}(x,\mu_{\mathrm{NLO}}^{2})=0.540(1-x)^{0.17}V_{\pi}(x,\mu_{\mathrm{NLO}}^{2}),
\end{equation}
\begin{equation}
V_{\bar{s}/K^{+}}(x,\mu_{\mathrm{NLO}}^{2})=V_{\bar{s}/K^{0}}(x,\mu_{\mathrm{NLO}}^{2})=V_{\pi}(x,\mu_{\mathrm{NLO}}^{2})-V_{u/K^{+}}(x,\mu_{\mathrm{NLO}}^{2}).
\end{equation}
Thus, we can calculate the asymmetries of light-flavor antiquark
distributions $\bar{d}(x)-\bar{u}(x)$ and strange-antistrange
quark distributions $S^{-}\equiv\int^{1}_{0}
x[s(x)-\bar{s}(x)]\textmd{d}x$ within the effective chiral quark
model. In this present work, we choose $m_{u}=m_{d}=330$~MeV,
$m_{s}=480$~MeV, $m_{\pi^{\pm}}=m_{\pi^{0}}=140$~MeV and
$m_{K^{+}}=m_{K^{0}}=495$~MeV. Constituent quark (CQ) model
distributions~\cite{hz81} and CTEQ6 parametrization~\cite{p02} as
two different kinds of constituent quark distributions of inputs
are adopted. The CQ model distributions have the following forms:
\begin{eqnarray}
u_{0}(x)&=&\frac{2}{\textmd{B}[c_{1}+1,c_{1}+c_{2}+2]}x^{c_{1}}(1-x)^{c_{1}+c_{2}+1},
\nonumber\\
d_{0}(x)&=&\frac{1}{\textmd{B}[c_{2}+1,2c_{1}+2]}x^{c_{2}}(1-x)^{2c_{1}+1},
\end{eqnarray}
where $B[i,j]$ is the Euler beta function and $c_{1}=0.65$ and
$c_{2}=0.35$ adopted from Ref.~\cite{hz81,kko99} are independent
of $Q^{2}$ with the number and momentum sum rules:
\begin{eqnarray}
\int^{1}_{0}u_{0}(x)\textmd{d}x=2, ~~~~
\int^{1}_{0}d_{0}(x)\textmd{d}x=1,\nonumber\\
\int_{0}^{1}xu_{0}(x)\textmd{d}x+\int_{0}^{1}xd_{0}(x)\textmd{d}x=1.
\end{eqnarray}
It is pointed out that there are other different values for
$c_{1}$  and $c_{2}$ suggested by Ref.~\cite{hy02}. And the CTEQ6
parameterizations are:
\begin{eqnarray}
u_{0}(x)&=&1.7199x^{-0.4474}(1-x)^{2.9009}\textmd{exp}[-2.3502x](1+\textmd{exp}[1.6123]x)^{1.5917},~~\nonumber\\
d_{0}(x)&=&1.4473x^{-0.3840}(1-x)^{4.9670}\textmd{exp}[-0.8408x](1+\textmd{exp}[0.4031]x)^{3.0000}.~~
\end{eqnarray}
As is well-known, the quark distributions measured by experiments
at certain $Q^{2}$ include not only non-perturbative intrinsic sea
but also perturbative extrinsic sea~\cite{bm96}. The intrinsic sea
quarks are relatively $Q^{2}$ independently (or less dependently)
multi-connected with the valence quarks in the nucleon, but the
extrinsic ones are generated mainly from the gluon splitting.
Because the correlations between the intrinsic and extrinsic sea
quarks are less $Q^{2}$ dependent, we give the distribution of
$\bar{d}(x)-\bar{u}(x)$ at low energy compared with the
experimental ones at high scale shown in Fig.~\ref{udx}. From the
figure, we find that the calculated results using the effective
chiral quark model match the experiments well with two different
sets~(CQ model and CTEQ parametrization) of inputs for the
constituent quark distributions. The difference between the
results from two different inputs may reflect some uncertainties
due to the evolution effect which should not cause a big influence
on the conclusion. Besides that, we also calculate the strange sea
distributions of $x(s(x)+\bar{s}(x))$ and $x(s(x)-\bar{s}(x))$
within this mechanism. For the distributions of
$x(s(x)+\bar{s}(x))$, we give the comparison of the results from
the effective chiral quark model calculations~(with two different
sets of inputs) and the NuTeV data parametrization~\cite{NUTEV}
obtained from the LO fits to the dimuon production cross section
in $\nu_{\mu}$Fe and $\bar{\nu}_{\mu}$Fe deep inelastic scattering
in Fig.~\ref{xsplussbar}. From the figure, we can find that the
results of the effective chiral quark model calculations are lower
than the parametrization of NuTeV data at arbitrary $x$, which may
be caused by the non-considered symmetric strange sea content
which will be discussed in Sec.~\ref{Sec:Four}

\begin{figure}
\begin{center}
\scalebox{0.98}{\includegraphics[0,16][320,230]{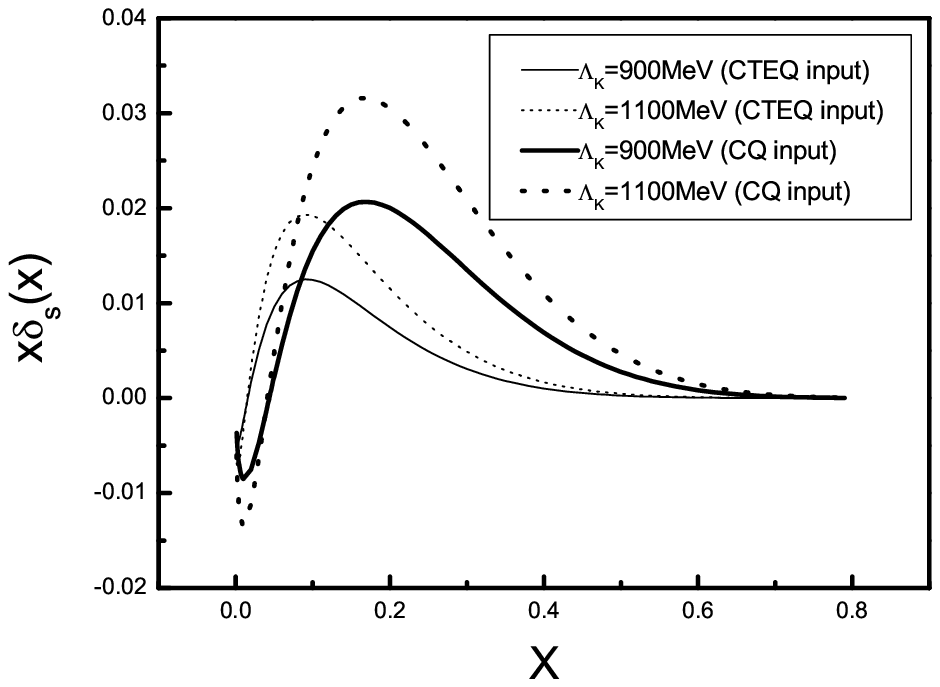}}
\caption{\small Distributions of $x\delta_{s}(x)$ with
$\delta_{s}(x)$=$s(x)-\bar{s}(x)$ in the effective chiral quark
model, when the inputs for $u_0(x)$ and $d_0(x)$ are both
constituent quark (CQ) model~(thick curves) and CTEQ6
parametrization~(thin curves) with $\Lambda_{K}=900$~MeV (solid
curves) and 1100~MeV (dotted curves ).}\label{ssbarx}
\end{center}
\end{figure}

In the effective chiral quark model, it should not be the same for
the momentum of strange quark ($s$) coming from the recoil
constituent quark and that of anti-strange quark ($\bar{s}$)
originating from the virtual GS bosons $K$.
Using Eqs.~(\ref{uddis}) and (\ref{udsdis}), we can obtain the
distributions for $x\delta_{s}(x)$ shown in Fig.~\ref{ssbarx} with
$\delta_{s}(x)=s(x)-\bar{s}(x)$. From the figure, we can see that
the magnitudes of $x\delta_{s}(x)$ with CQ model as input are
almost twice larger than those with CTEQ parametrization as input,
however, the contributions to the NuTeV anomaly, which will be
shown in the next section, are similar.


\section{Application to the NuTeV anomaly}
In 1973, Paschos and Wolfenstein~\cite{pash73} suggested to
consider the observable Eq.~(\ref{ratio}) by measuring the ratio
of neutral-current to charged-current cross sections for neutrino
and antineutrino on isoscalar target to reduce the uncertainties
from the productions of the heavy flavor quarks. In
Eq.~(\ref{ratio}), $\sigma^{\nu N}_{NC}$~($\sigma^{\bar{\nu}
N}_{NC}$) is the integral of neutral-current inclusive
differential cross section for neutrino~(antineutrino) over $x$
and $y$, and the same to $\sigma^{\nu N}_{CC}$~($\sigma^{\bar{\nu}
N}_{CC}$). The NuTeV Collaboration was inspired by the P-W
relation to measure the Weinberg angle $\sin^{2}\theta_{w}$ using
high statistics separated neutrino and antineutrino beams which
arise from $\pi$ and $K$ mesons decay following the interaction of
800~GeV protons at FNAL. Because  the charged-current events
contain a final muon that penetrates substantially farther than
the hadron shower, neutral-~and charged-current events are
distinguished by the event length in the counter. The resulting
interaction events must be deposited between 20~GeV and 180~GeV in
the calorimeter and will be found in the NuTeV detector. The most
general form of the differential cross section for neutral-current
interactions initialed by (anti-)neutrino on nucleon target
is~\cite{lt98}:
\begin{eqnarray}
\frac{\textmd{d}^{2}\sigma^{\nu(\bar{\nu})}_{NC}}{\textmd{d}
x\textmd{d} y}&=&\pi
s\left(\frac{\alpha}{2\sin^{2}\theta_{w}\cos^{2}\theta_{w}M^{2}_{Z}}
\right)^{2}(\frac{M^{2}_{Z}}{M^{2}_{Z}+Q^{2}})^{2} [xyF^{Z}_{1}(x,Q^{2}) \nonumber\\
&&
+(1-y-\frac{xym^{2}_{N}}{s})F^{Z}_{2}(x,Q^{2})\pm(y-\frac{y^{2}}{2})xF^{Z}_{3}(x,Q^{2})].
\end{eqnarray}
And similarly, for the (anti-)neutrino-nucleon charged-current
reaction, the differential cross section has the form~\cite{lt98}:
\begin{eqnarray}
\frac{\textmd{d}^{2}\sigma^{\nu(\bar{\nu})}_{CC}}{\textmd{d} x
\textmd{d} y}&=&{\pi}s\left(\frac{\alpha}{2\sin^{2}\theta_{w}
M^{2}_{W}}\right)^{2}(\frac{M^{2}_{W}}{M^{2}_{W}+Q^{2}})^{2}
 [xyF^{W^{\pm}}_{1}(x,Q^{2}) \nonumber\\
&& +(1-y-\frac{xym^{2}_{N}}{s})
F^{W^{\pm}}_{2}(x,Q^{2})\pm(y-\frac{y^{2}}{2})xF^{W^{\pm}}_{3}(x,Q^{2})
],
\end{eqnarray}
where $M_{Z}$ and $M_{W}$ are the masses of the neutral-  and
charged-current interacting weak vector bosons, respectively,
$\theta_{w}$ is the Weinberg angle, $x=Q^{2}/2p\cdot q$, $y=p\cdot
q/p\cdot k$, and the square of the four momentum transfer for the
reaction is $Q^{2}=-q^{2}$, $k$ ($p$) is the momentum of initial
state for neutrino or antineutrino (nucleon) and $s=(k+p)^{2}$.
Besides these, $F^{Z(W^{\pm}) p}_{i}(x,Q^{2})$ are the structure
functions on proton ($p$), which only depend on $x$ as
$Q^{2}\rightarrow\infty$, and are written in terms of the parton
distributions as~\cite{lt98}
\begin{eqnarray}
\lim_{Q^{2}\rightarrow\infty}F^{Zp}_{1}(x,Q^{2})&=&\frac{1}{2}\left[\left((f^{u}_{V})^{2}+(f^{u}_{A})^{2}\right)\left(u^{p}(x)+\bar{u}^{p}(x)
    +c^{p}(x)+\bar{c}^{p}(x)\right)\right.\nonumber\\&&\left.+\left((f^{d}_{V})^{2}+(f^{d}_{A})^{2}\right)\left(d^{p}(x)+\bar{d}^{p}(x)
    +s^{p}(x)+\bar{s}^{p}(x)\right)\right],\nonumber\\
\lim_{Q^{2}\rightarrow\infty}F^{Zp}_{3}(x,Q^{2})&=&2\left[f^{u}_{V}f^{u}_{A}\left(u^{p}(x)-
    \bar{u}^{p}(x)+c^{p}(x)-\bar{c}^{p}(x)\right)\right.\nonumber\\&&\left.+f^{d}_{V}f^{d}_{A}\left(d^{p}(x)-
    \bar{d}^{p}(x)+s^{p}(x)-\bar{s}^{p}(x)\right)\right]. \nonumber\\
F^{Zp}_{2}(x,Q^{2})&=&2xF^{Zp}_{1}(x,Q^{2}).\label{struc}
\end{eqnarray}
And the structure functions of charged-current have forms:
\begin{eqnarray}
\lim_{Q^{2}\rightarrow\infty}F^{W^{+}p}_{1}(x,Q^{2})&=&d^{p}(x)+\bar{u}^{p}(x)+s^{p}(x)+\bar{c}^{p}(x),\nonumber\\
\lim_{Q^{2}\rightarrow\infty}F^{W^{-}p}_{1}(x,Q^{2})&=&u^{p}(x)+\bar{d}^{p}(x)+\bar{s}^{p}(x)+c^{p}(x),\nonumber\\
\frac{1}{2}\lim_{Q^{2}\rightarrow\infty}F^{W^{+}p}_{3}(x,Q^{2})&=&d^{p}(x)-\bar{u}^{p}(x)+s^{p}(x)-\bar{c}^{p}(x),\nonumber\\
\frac{1}{2}\lim_{Q^{2}\rightarrow\infty}F^{W^{-}p}_{3}(x,Q^{2})&=&u^{p}(x)-\bar{d}^{p}(x)-\bar{s}^{p}(x)+c^{p}(x),\nonumber\\
F^{W^{\pm}p}_{2}(x,Q^{2})&=&2xF^{W^{\pm}p}_{1}(x,Q^{2}).\label{stru}
\end{eqnarray}
One can obtain the structure functions for neutron~$n$ by
replacing the superscripts $p\rightarrow$$n$ everywhere in
Eqs.~(\ref{struc}) and~(\ref{stru}). In Eq.~(\ref{struc}),
$f^{u}_{V}$, $f^{u}_{A}$, $f^{d}_{V}$ and $f^{d}_{A}$ are vector
and axial-vector couplings:
$$f^{u}_{V}=\frac{1}{2}-\frac{4}{3}\sin^{2}\theta_{w},  \ \ f^{u}_{A}=\frac{1}{2},$$
$$f^{d}_{V}=-\frac{1}{2}+\frac{2}{3}\sin^{2}\theta_{w}, \ \ f^{d}_{A}=-\frac{1}{2}.$$ When the charge symmetry
\begin{eqnarray}
   d^{n}(x)&=&u^{p}(x),\nonumber\\
   u^{n}(x)&=&d^{p}(x),\nonumber\\
  s^{n}(x)&=&s^{p}(x)=s(x),\nonumber\\
  c^{n}(x)&=&c^{p}(x)=c(x), \label{pdis}
\end{eqnarray}
and $c(x)=\bar{c}(x)$ are valid, the structure functions for the
isoscalar target with the $s(x)\neq\bar{s}(x)$ are given by
\begin{eqnarray}
\lim_{Q^{2}\rightarrow\infty}F^{W^{+}N_{0}}_{1}(x,Q^{2})&=&\frac{1}{2}
[d^{p}(x)+\bar{d}^{p}(x)+u^{p}(x)+\bar{u}^{p}(x)+2s(x)+2c(x)],\nonumber\\
\lim_{Q^{2}\rightarrow\infty}F^{W^{-}N_{0}}_{1}(x,Q^{2})&=&\frac{1}{2}
[d^{p}(x)+\bar{d}^{p}(x)+u^{p}(x)+\bar{u}^{p}(x)+2\bar{s}(x)+2c(x)],\nonumber\\
\lim_{Q^{2}\rightarrow\infty}F^{W^{+}N_{0}}_{3}(x,Q^{2})&=&
d^{p}(x)+u^{p}(x)-\bar{u}^{p}(x)-\bar{d}^{p}(x)+2s(x)-2c(x),\nonumber\\
\lim_{Q^{2}\rightarrow\infty}F^{W^{-}N_{0}}_{3}(x,Q^{2})&=&
d^{p}(x)+u^{p}(x)-\bar{u}^{p}(x)-\bar{d}^{p}(x)-2\bar{s}(x)+2c(x),\nonumber\\
F^{W^{\pm}N_{0}}_{2}(x,Q^{2})&=&2xF^{W^{\pm}N_{0}}_{1}(x,Q^{2}),\label{str}
\end{eqnarray}
where
$F^{W^{+}N_{0}}_{i}(x,Q^{2})=\frac{1}{2}(F^{W^{+}p}_{i}(x,Q^{2})+F^{W^{+}n}_{i}(x,Q^{2}))$.
From above equations, we obtain the modified P-W relation with the
asymmetry of $s(x)$ and $\bar{s}(x)$ considered:
\begin{eqnarray}
  R^{-}_{N}=\frac{\sigma^{\nu N}_{NC}-\sigma^{\bar{\nu}N}_{NC}}{\sigma^{\nu N}_{CC}-\sigma^{\bar{\nu}N}_{CC}}
= R^{-}-\delta R^{-}_{s},\label{correction}
\end{eqnarray}
where $\delta R^{-}_{s}$ is the correction brought by the
asymmetric distribution of strange-antistrange sea and  $R^{-}$ is
the naive P-W relation with
\begin{eqnarray}
  \delta
  R^{-}_{s}=(1-\frac{7}{3}\sin^{2}\theta_{w})\frac{S^{-}}{Q_V+3 S^{-}},\label{rs}
\end{eqnarray}
and we define $Q_V \equiv\int^{1}_{0}
x[u_{V}(x)+d_{V}(x)]\textmd{d}x$ and
$S^-\equiv\int^{1}_{0}x(s(x)-\bar{s}(x))\textmd{d}x$. From above
equations, we can find that the asymmetry of $s(x)$ and
$\bar{s}(x)$ distributions should bring correction to
$\sin^{2}\theta_{w}$. The results are given in
table~\ref{cq}~and~\ref{cteq} with the values of $\delta
R^{-}_{s}$ ranging from 0.00297~(0.00312) to 0.00826~(0.00867)
corresponding to $\Lambda_{K}=\Lambda_{\pi}=900-1500$~MeV for
different inputting parameters. However, the $SU_{f}(3)$ symmetry
breaking requires smaller $\langle P_{K}\rangle$ and $\langle
P_{\eta}\rangle$ \cite{song}, so that we should adopt a smaller
value for $\Lambda_{K}$ such as from 900~MeV to 1100~MeV. Thus,
the correction brought by the asymmetry of $s(x)$ and $\bar{s}(x)$
distributions should remove the NuTeV anomaly by about 60-100\%
within the effective chiral quark model. Although the magnitude of
$S^{-}$ with CQ model as input is almost twice larger than that
with CTEQ parametrization as input, as pointed out in
Sec.~\ref{Sec:Two}, the calculated results of $\delta R^{-}_{s}$
given in Table~\ref{cq}~and~\ref{cteq} are similar and not
sensitive to the different inputs at fixed $\Lambda_{K}$. The
reason may be  that the uncertainties in the numerator and
denominator of the Eq.~(\ref{rs}) brought by the model can cancel
out each other.
\begin{table}[!htbp]\caption {\label{cq}The results for  CQ model input}
\begin{center}
\begin{tabular}{|c|c|c|c|c|}
  \hline
  \hline
  $\Lambda_{k}$ & Z & $Q_{v}$ & $S^{-}$ & $\delta R_{s}^{-}$ \\
  \hline
  900 & 0.74888 & 0.86376 & 0.00558 & 0.00297 \\
  1000 & 0.73996 & 0.85484 & 0.007183 & 0.00384 \\
  1100 & 0.73063 & 0.84551 & 0.00879 & 0.00473 \\
  1200 & 0.72107 & 0.83595 & 0.01040 & 0.00562 \\
  1300 & 0.71143 & 0.82632 & 0.01198 & 0.00651 \\
  1400 & 0.70181 & 0.81669 & 0.01353 & 0.00740 \\
  1500 & 0.69227 & 0.80715 & 0.01503 & 0.00826 \\
    \hline
\end{tabular}
\end{center}
\end{table}
\begin{table}[!htbp]\caption {\label{cteq}The results for  CTEQ parametrization input}
\begin{center}
\begin{tabular}{|c|c|c|c|c|}
  \hline
  \hline
  $\Lambda_{k}$ & Z & $Q_{v}$ & $S^{-}$ & $\delta R_{s}^{-}$ \\
  \hline
 900 & 0.74888 & 0.37089 & 0.00252 & 0.00312 \\
  1000 & 0.73996 & 0.36686 & 0.00322 & 0.00402 \\
  1100 & 0.73063 & 0.36247 & 0.00398 & 0.00498 \\
  1200 & 0.72107 & 0.35831 & 0.00468 & 0.00590 \\
  1300 & 0.71143 & 0.35395 & 0.00539 & 0.00683 \\
  1400 & 0.70181 & 0.34960 & 0.00612 & 0.00780 \\
  1500 & 0.69227 & 0.34528 & 0.00675 & 0.00867 \\
    \hline
\end{tabular}
\end{center}
\end{table}

\section{Discussion on Additional Symmetric Sea Contributions
\label{Sec:Four}} 
\begin{figure}
\begin{center}
\scalebox{0.98}{\includegraphics[0,16][310,230]{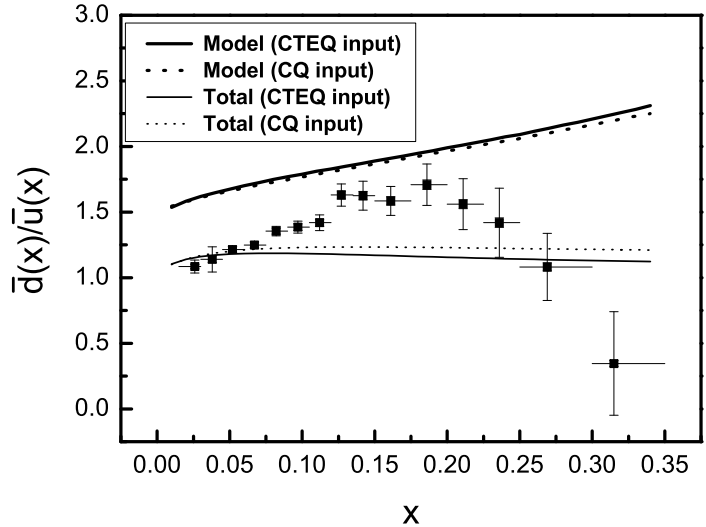}}
\caption{\small The distributions of $\bar{d}(x)/\bar{u}(x)$ for
$\Lambda_{\pi}=1500$~MeV. The thick solid and dotted curves are
the chiral quark model results with CTEQ6 parametrization and the
CQ model as inputs. The thin curves are the corresponding results
with both chiral quark model results and the symmetric sea
contribution $\delta\bar{u}(x)=\delta\bar{d}(x)$ estimated by
difference between the NuTeV data parametrization and the chiral
quark model results. And the experimental data come from
E866/NuSea~($Q^{2}=54$~GeV$^{2}$) experiments~\cite{e01}.
}\label{dubarr}
\end{center}
\end{figure}

\begin{figure}
\begin{center}
\scalebox{1.4}{\includegraphics[0,16][310,235]{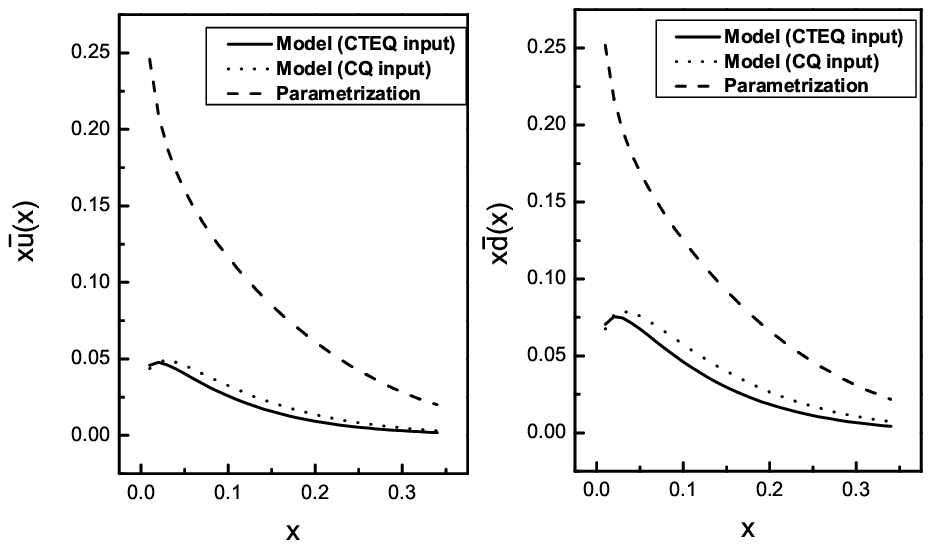}}
\caption{\small The distributions of $\bar{d}(x)$ and $\bar{u}(x)$
for $\Lambda_{\pi}=1500$~MeV. The solid and dotted curves are the
effective chiral quark model results with CTEQ6 parametrization
and CQ model inputs, and the dashed curve is for the NuTeV data
parametrization.} \label{xdubarr}
\end{center}
\end{figure}

The asymmetries of light-flavor antiquark and
strange-antistrange distributions originated from the
nonperturbative sea have been discussed within the effective
chiral quark model. Although the asymmetries of light-flavor
antiquark and strange sea content in the nucleon are mainly coming
from the intrinsic sea, the extrinsic sea should be also
considered from a strict sense when we investigate the
distributions of $\bar{d}(x)-\bar{u}(x)$, $\bar{d}(x)/\bar{u}(x)$
and $s(x)/\bar{s}(x)$. From the calculations, we find that the
effective chiral quark model predictions of
$\bar{d}(x)-\bar{u}(x)$ distributions keep consistence with the
experimental data, but the predictions of $\bar{d}(x)/\bar{u}(x)$
distributions are not good in matching the experimental data at
small and large $x$ by using the effective chiral quark model
before taking into account additional contribution from the
symmetric sea~(e.g. $\delta\bar{d}(x)=\delta\bar{u}(x)$).
Because the contribution of the symmetric sea
effect~($\delta\bar{d}(x)=\delta\bar{u}(x)$) cancels out each
other in the distributions of $\bar{d}(x)-\bar{u}(x)$, the
distributions of $\bar{d}(x)-\bar{u}(x)$
match the experimental data well. For the distributions of
$\bar{d}(x)/\bar{u}(x)$, when the symmetric sea content is
considered, the distribution of $\bar{d}(x)/\bar{u}(x)$ becomes:
\begin{eqnarray}
\frac{\bar{d}(x)}{\bar{u}(x)}&=&\frac{\bar{d}(x)+\delta\bar{d}(x)-\bar{u}(x)-\delta\bar{u}(x)}
{\bar{u}(x)+\delta\bar{u}(x)}+1\nonumber\\
&=&\frac{\bar{d}(x)-\bar{u}(x)}{\bar{u}(x)+\delta\bar{u}(x)}+1.\label{duratio}
\end{eqnarray}
From the Eq.~(\ref{duratio}), we can know that the ratio of
$\bar{d}(x)/\bar{u}(x)$ will decrease if the symmetric sea is not
zero.
Here, we estimate the symmetric sea contribution
$\delta\bar{u}(x)+\delta\bar{d}(x)$
($\delta\bar{u}(x)=\delta\bar{d}(x)$) by the difference between
the parametrization~\cite{cteq4} of $\bar{u}(x)+\bar{d}(x)$ and
the results predicted by effective chiral quark model. Thus, the
distributions of $\bar{d}(x)/\bar{u}(x)$ without and with
symmetric sea contribution are obtained and shown in
Fig.~\ref{dubarr}. From the figure, we can find that a large
suppression occurs by including the symmetric sea contribution,
which implies that the effective chiral quark model only provides
a fraction of the total light-flavor in the nucleon, and that a
significant fraction of symmetric sea content is required. This
also can be found in Fig.~\ref{xdubarr}, which displays the
comparison of the distributions for $\bar{u}(x)$ and $\bar{d}(x)$
obtained from the parametrization~\cite{cteq4} and the effective
chiral quark model calculations. The role of the symmetric sea is
similar to that by including some isoscalar~(pseudoscalar) meson
$\sigma$~($\eta$) contribution in the meson cloud model, as
pointed in Ref.~\cite{huang04}. However, the effective way of
taking into account the symmetric sea contribution is not good in
reproducing the shape of $\bar{d}(x)/\bar{u}(x)$, and more work
both theoretically and experimentally should be done along this
direction.


\begin{figure}[htbp]
\begin{center}
\scalebox{1.2}{\includegraphics[0,16][310,250]{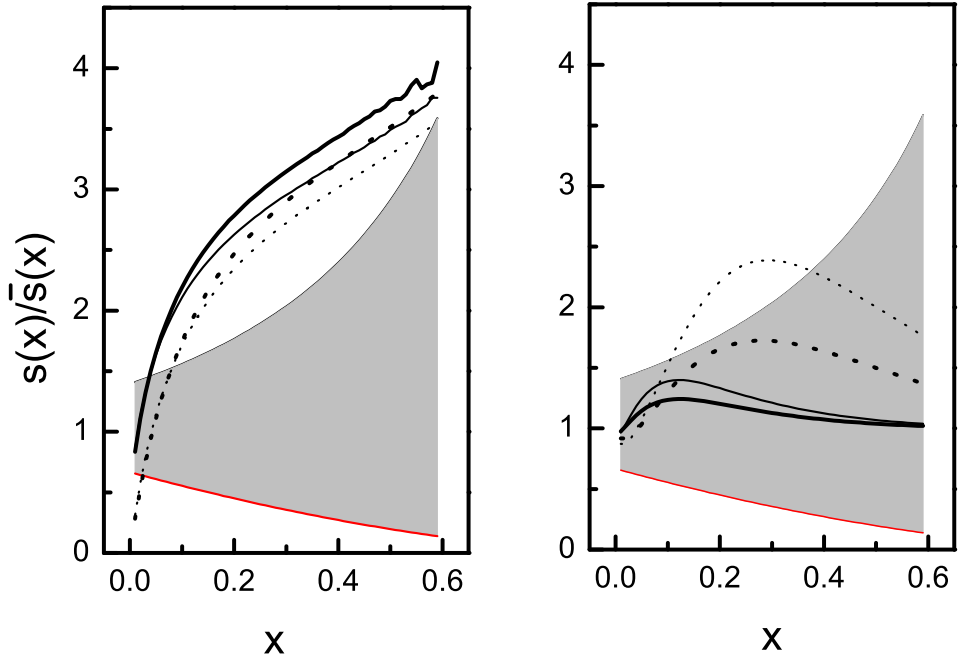}}
\end{center}
\caption[*]{\small Distributions of $s(x)/\bar{s}(x)$, where the
shadowing area is the error range of NuTeV. The thick and thin
solid curves are the effective chiral quark model results for
$s(x)/\bar{s}(x)$ when $\Lambda_{K}=900$ and $1100$~MeV with CTEQ6
parametrization as input. The thick and thin dotted curves are the
effective chiral quark model results for $s(x)/\bar{s}(x)$ when
$\Lambda_{K}=900$ and $1100$~MeV with CQ model as input. The left
side is the prediction by the effective chiral quark model only
and the right side is the result by including both the prediction
of the effective chiral quark model and the symmetric sea
contribution estimated by the difference between the NuTeV data
parametrization and the effective chiral quark model
result.}\label{ssbar}
\end{figure}

Within the effective chiral quark model, the strange-antistrange
asymmetry mainly comes from the $K$ meson clouds which accompany
the constituent quarks. From the calculations we know that the
corrections coming from the strange asymmetry with different
inputs of constituent quark distributions are large enough to
explain the NuTeV anomaly. It is also necessary to mention that
the ratio of $s(x)/\bar{s}(x)$ in the model does not necessarily
conflict with the experimental data in fact,
because the experimental data include the effect of the intrinsic
and extrinsic sea.
In this paper, we calculate the ratio of $s(x)/\bar{s}(x)$ without
and with symmetric strange sea contribution, where the symmetric
sea contribution is estimated effectively by the difference
between  $s(x)+\bar s (x)$ from the NuTeV data parametrization and
that in the effective chiral quark model calculation, in similar
to what we estimated above for the symmetric light-flavor sea
quarks. The so called ``NuTeV data parametrization" here is
obtained by the NuTeV collaboration from the results of the
leading order fits to the cross-section extracted from the NuTeV
experimental data~\cite{NUTEV}. The strange sea parameterizations
which break $\int_0^1 s(x) {\mathrm{d}} x=\int_0^1
\bar{s}(x){\mathrm{d}} x$ as pointed out as in~\cite{bm96,0405037}
are defined conventionally by
\begin{eqnarray}
s(x)=\kappa\frac{\bar{u}(x)+\bar{d}(x)}{2}(1-x)^{\alpha},\nonumber\\
\bar{s}(x)=\bar{\kappa}\frac{\bar{u}(x)+\bar{d}(x)}{2}(1-x)^{\bar{\alpha}}.
\end{eqnarray}
where the values of $\kappa$, $\bar{\kappa}$, $\alpha$,
$\bar{\alpha}$ are taken from the NuTeV set of the Table III in
Ref.~\cite{NUTEV}. And we give the comparison of $s(x)/\bar s(x)$
from the model prediction and that from the dimuon
measurements~\cite{NUTEV} in Fig.~\ref{ssbar}.
From the left one of the figure without the symmetric sea content,
we can see that the ratios are out of the area for errors.
However, the distributions of $s(x)/\bar{s}(x)$  including the
symmetric sea contribution are almost within the range of the
experimental errors in the right one. This case is similar to the
$\bar{d}(x)/\bar{u}(x)$ distribution and indicates that the
effective chiral quark model only provides a fraction of the
strange sea content, and that a significant fraction of symmetric
strange sea content is still needed. Fortunately, the strange
asymmetry  of the effective chiral quark model can cause a large
reduction of the NuTeV anomaly, although the asymmetric strange
sea is only a small part of the total strange sea content of the
nucleon.

\section{Summary}

In this work, we presented the calculations of the asymmetries of
light-flavor sea quark distributions $\bar{d}(x)-\bar{u}(x)$ and
strange-antistrange $s(x)-\bar{s}(x)$ within the effective chiral
quark model with more details along with our previous
work~\cite{dxm04} by using two different sets of parameterizations
as inputs and find that the results for $\bar{d}(x)-\bar{u}(x)$
match the experimental measurements well. The distributions of
$\bar{d}(x)/\bar{u}(x)$ do not match with the experimental data,
but there is a large suppression of the ratio when the symmetric
sea contribution is considered. The contributions of symmetric sea
effects are estimated effectively by the difference between the
effective chiral quark model results and the data parametrization
results. Fig.~\ref{xsplussbar} and Fig.~\ref{xdubarr} indicate
that the predictions of the effective chiral quark model provide
only a fraction of the total light-flavor sea quarks and strange
sea content of the nucleon. Also, we point out that the calculated
results for $s(x)/\bar{s}(x)$ in the effective chiral quark model
are consistent with the available experiments with an additional
symmetric sea
contribution being included effectively. 
More noticeably, the asymmetry of strange-antistrange
distributions can bring a significant contribution to the NuTeV
defect by at least 60\% with reasonable parameters within this
model, and the results are not sensitive to the different inputs
of constituent quark distributions, although the intrinsic sea is
not the dominant part in the nucleon sea.
The effective chiral quark model is thus successful in explaining
the nucleon sea anomalies not only about the GSR violation and the
proton spin problem, but also about the NuTeV defect. Also, this
may imply that the NuTeV anomaly can be considered as a
phenomenological support to the strange-antistrange asymmetry of
the nucleon sea. So it is more important that more precision
experiments should be carried out to enable direct and more
accurate determination of strange-antistrange sea quark
distributions in the future.

\section{Acknowledgments}

This work is partially supported by National Natural Science
Foundation of China (Nos.~10025523, 90103007, and 10421003), and
by the Key Grant Project of Chinese Ministry of Education
(No.~305001).

\end{document}